# Surface plasmon assisted electron pair formation in strong electromagnetic field

N. Kroó[1], P. Rácz[1(a), 2] AND S. Varró[1]

[1] Wigner Research Centre for Physics, Konkoly-Thege M. út 29-33, 1121 Budapest, Hungary
[2] ELI-HU Nonprofit Kft. Dugonics Tér 13, 6720 Szeged, Hungary

**Abstract**

The basis of low-temperature superconductivity has been set to be the pair formation of electrons, due to their effective attraction. The appearance of an effective attraction potential has also been predicted for electron-electron scattering in the presence of strong, inhomogeneous radiation field. In the present work the strong electromagnetic fields were created by femtosecond Ti:Sa lasers, used to excite surface plasmons in gold films at room temperature in the Kretschmann geometry. Experimental investigations were carried out using a surface plasmon near field scanning tunneling microscope, measuring it's response to the excitation at hot spots on the gold surface. Furthermore, the spectra of photoelectrons, liberated by multi-plasmon absorption, has also been measured by a time-of-flight spectrometer. In both cases new type of anomalies in the electron signal have been measured in the same intensity range, whose existence may be qualitatively understood, by using the intensity-dependent expression for the effective electron-electron scattering potential, derived earlier in a different context.

Room temperature superconductivity has been our dream for a long time. Some ideas can be found on this issue e.g. in [1] where the possibility to find this phenomenon even in living tissues has also been discussed. The first question raised in this reference, as in many other papers elsewhere, is electron pairing, where the mediators of pairing are not phonons, but other quasi-particles.

The appearance of an effective attraction potential between the electrons has also been studied theoretically for a completely different case, namely for electron-electron interaction in strong laser fields in vacuum, by one of the authors of this paper in [2]. Motivated by the results of this earlier study, our attempt to tackle this problem has been to perform two types of experiments, in order to see if surface plasmons could be the mediators of electron pairing at room temperature.

A scanning tunneling microscope (STM) has been used for the first set of experiments since STM-s have been widely used for many years to study the near field of surface plasmon oscillations (SPO) [3]. It has been found e.g. by one of the authors, that in addition to the usual thermal response of the STM to SPO excitation (slow signals) in some, so-called hot spots [4,5] of the surface, where SPO-s get localized, fast signals are detected. In these cases the response of the STM to the SPO excitation is practically determined by the time constant of the microscope, being typically in the 10 μs range. It has been found [6], that this rectified signals might be negative even if the STM is positively biased.

In most of our experiments a ~30 μs window of a CW semiconductor laser has been used as the SPO exciting source, but it has been found, that femtosecond Ti:Sa laser pulses [7,8] might also be used to excite SPO-s and they can be detected by the STM. It has also been found, that these rectified signals do not disappear when the bias voltage on the STM is zero [9].

The present paper reports some of our findings for this case, when the SPO exciting laser intensity has been in the 10-220 GW/cm² range. These intensity values are orders of magnitude lower, than those needed to induce similar nonlinear optical effects [10] in free space. This difference is due to the EM field enhancement effect of the SPO-s. The negative STM response to laser excitation at these spots has been studied at zero STM bias.

The measurements were performed with a long cavity Ti:Sa laser oscillator (~ 80 m), delivering 100-120 fs, 0.1 µJ laser pulses at 800 nm central wavelength with 3.6 MHz repetition rate. From this pulse train 1 to 5 pulses have been selected with 1.2 kHz repetition rate by a pulse picker based on the Pockels effect. SPOs were generated in the Kretschmann configuration on a right-angle fused silica prism on which a 45 nm Au layer was evaporated for the optimal SPO coupling. (Laser pulses were focused by an f = 3 cm lens onto the gold film.) The SPO (near field) excitations were investigated by the STM [11]. During the measurement between the tip (W) and the metal layer a bias voltage was applied. The tunneling current was processed by a low-noise current-to-voltage converter and amplified by a factor of $10^9$. The electronic signal of the amplifier was recorded by an A/D converter at a sampling rate of 500 kHz. At certain positions of the gold film (hot spots) the SPO excitations produced fast signals [5] which were recorded by the STM as a function of the incident laser intensity and bias voltage. For the bias voltage dependence measurements, the surface scanning was stopped and the sample-to-tip distance was kept constant by turning the feedback loop off during the measurement. At each bias voltage 30 measurements were averaged. Fig.1 shows a typical result of our measurements, where the integral of the recorded STM signal as the function of laser intensity at zero STM bias voltage is plotted.

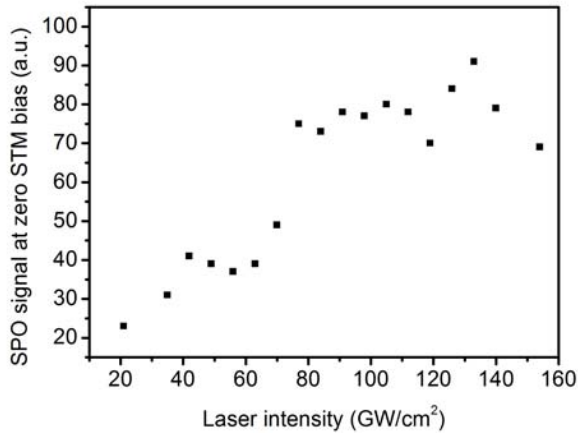

Fig.1: A typical laser intensity dependence plot of the STM response to SPO excitation in the Kretschmann geometry on a 45 nm gold film by 100 fs 800 nm Ti:Sa laser pulses

It has been found to our greatest surprise, that the laser intensity dependence of this SPO signal is not monotonous, but has a step-like character as shown in Fig. 1. What is, however, even more surprising is the change of the decay time of the SPO signal when changing the laser intensity. Although the length of the laser pulse has been 100-120 fs and the life-time of SPO-s is also in the 100 fs range at this laser wavelength, the decay time of this signal has been broadened with tens of microseconds with increasing laser intensity up to about 70-80 GW/cm², but decreasing with about the same rate, when the intensity has been further increased. This is shown in Fig. 2 for measurements at one hot-spot on the gold surface but reproduced at several other hot spots too.

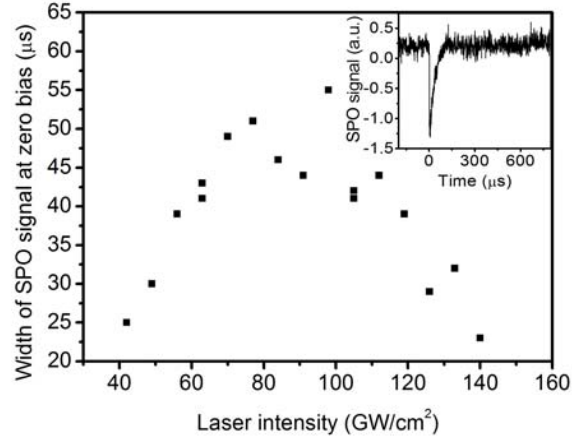

Fig. 2: Laser intensity dependence of the temporal response of the STM (the width of the pulse at 1/e height) to 100fs 800nm Ti:Sa laser pulse excitation in gold. The width of the (not broadened) signal is ~15μs. The differently shaped dots represent measurements at different hot spots on the gold surface

In order to find additional checks for the existence of an anomaly, a second set of experiments has also been carried out. Surface plasmon induced electron emission and acceleration experiments have been studied with a Ti:Sa based chirped-pulse amplifier-laser system (Coherent Legend Elite HE-USP). The amplifier delivers 40 fs pulses at 800 nm central wavelength with 1 kHz repetition rate and with 4 mJ pulse energy. The pulse energy was reduced to below 1 μJ (and was focused by an f =20 cm lens to a 20 μm spot on the gold surface). The surface plasmon oscillations were generated also in the Kretschmann configuration and the right-angle fused silica prism with the gold film was used as the window for the vacuum chamber (the pressure in the chamber was around $10^{-7}$ mbar). The Au layer was evaporated onto the vacuum side of this prism. The emitted electrons were investigated by a time-of-flight electron spectrometer (Kaesdorf Electron TOF) in this vacuum chamber. The time of flight spectra were recorded by a high temporal resolution multiscaler card (P7889) with 100 ps resolution. Due to the relatively large surface area where the plasmons were excited, our measurements represent the average of numerous hot spots and the emission from spots with much lower localized plasmon fields.
The acceptance cone of TOF was relative high with an applied accelerating voltage inside the TOF spectrometer. Therefore it was possible to measure the total photocurrent as a function of the incident laser intensity as well as the spectrum of electrons, simultaneously.

One typical measurement of the spectrum of emitted electrons is shown in Fig. 3 where the parameter at each curve is the actual laser intensity in GW/cm² units. The observation of higher emitted electron energies has been reported in several earlier papers, including some of our earlier results [12]. Since the work function of gold is 4.6 eV and the photon energy of our laser is 1.5 eV, four photons (plasmons) are needed to get an electron out from the metal surface. Therefore in the intensity range, where the perturbation theory may be applied linear intensity dependence of the total electron number is expected in a double logarithmic plot with the slope of 4. This has been experimentally found as seen in Fig. 4.

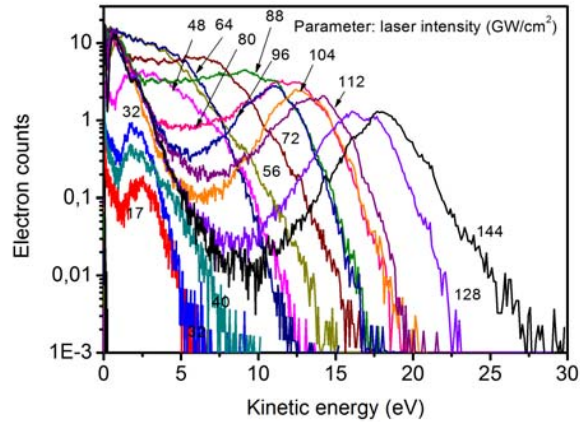

Fig. 3: SPO mediated electron emission spectra of a 45nm thick gold film as the function of the exciting laser intensity. The Ti:Sa pulsed laser light (40fs, 800nm, 1kHz) intensity is given in GW/cm² units and is the parameter on each curve. The spectra were measured by a time-of-flight spectrometer and converted to energy scale.

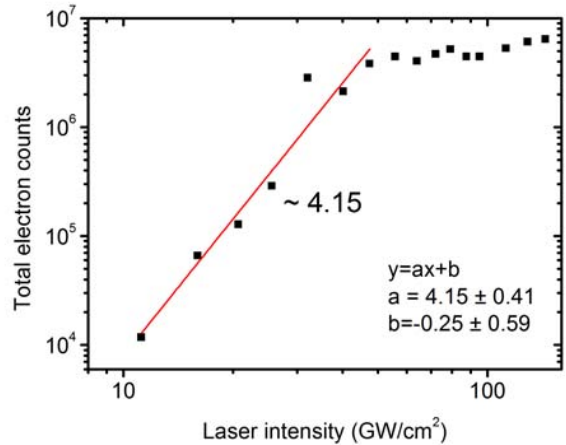

Fig. 4: Laser intensity dependence of the total number of electrons emitted via multiplasmon mediated excitation from $2 \cdot 10^5$ shots of the Ti:Sa laser.

But it is also seen in Fig. 3, that the electron spectrum is double peaked. If we plot the total number of electrons as a function of the laser intensity for the lower and higher energy peaks, the lower one has the slope of 3 in a double logarithmic plot before "saturation" around 80 GW/cm². This may represent surface regions, where the work function of gold is lower than 4.5 and therefore 3 plasmons are enough to mediate the emission of one electron. The higher one, however has a higher than four (~6) slope in the lower intensity range, but similarly as in the STM case around 70-80 GW/cm² it changes to a negative slope, and the count number has a 3 orders of magnitude dynamics as shown in Fig. 5, where the results of one of our measurements at a hot spot on the gold surface are presented. This figure practically shows the same dependence as that of Fig. 2, where, however, the change in decay time is only about 5-fold. The extremely long decay time of the SPO signals in the laser intensity region around 80 GW/cm² (Fig. 2), compared to the ~8 orders of magnitude shorter excitation time is the indication of a process allowing the electron tunneling only with a significant delay. That is why we decided to conduct this second set of plasmon experiments, in the hope to find further evidence and better understanding of the effect found with the STM, and seemingly we succeeded to get it.

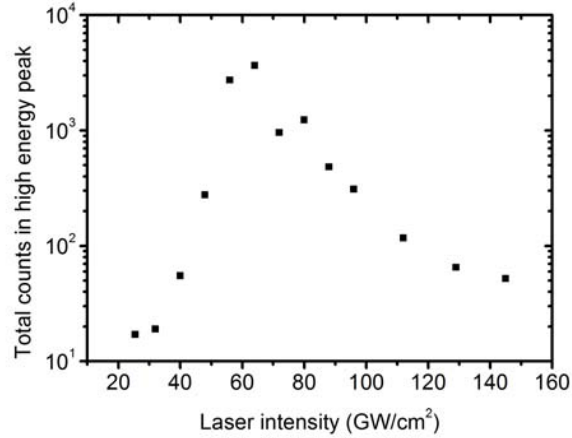

Fig. 5: Laser intensity dependence of the total number of electrons emitted into the higher energy peak of the electron spectra of Fig. 3. The differently shaped dots represent measurements at different spots of the gold surface.

An other interesting observation in these measurements is, that the TOF spectra have a tail, where only electrons can be detected, leaving the gold surface significantly later than the plasmon lifetime. The plot of the total number of these electrons as function of the laser intensity is shown in Fig. 6. The maximum of this plot is correlated with the minimum of Fig. 5, in agreement with our expectations. This finding is at the same time the qualitative bridge between the STM and TOF measurements.

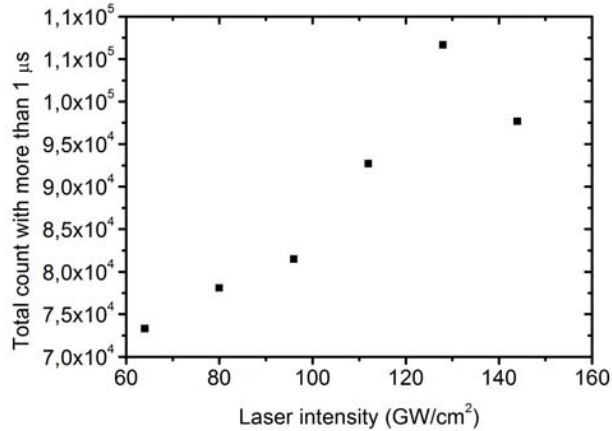

Fig. 6: Laser intensity dependence of the total number of electrons emitted from the gold surface more than 1μs after the laser pulse, i.e. with significant delay.

When looking for some qualitative hints towards the explanation of these observations we refer to the paper of one of the authors [2], where electron-electron scattering in vacuum and high-intensity laser fields has been studied theoretically. The resulting effective interaction potentials associated to different multiphoton channels contain a Bessel function, which may take on both positive and negative values, and in the negative regions an effective attraction results, which may lead to electron pairing. According to Eq. (2.18), and the consideration of ter Eq. (3.4) of Ref [2], the effective scattering potential has been given as

$$V_{eff}(\mathbf{r}) = V(\mathbf{r})J_n[z_1 \sin(\mathbf{k}\cdot\mathbf{r}/2)], \, z_1 = 2\mu(c\Delta p'_\perp/\hbar\omega), \, \mu = eF/mc\omega, \qquad (1)$$

where $V(\mathbf{r}) = e^2/r$ represents the original Coulomb repulsion of the electrons, with $r$ being their relative distance, $\mathbf{k}$ is the wave-vector of the surface plasmon, and $\Delta p'_\perp$ is the modulus of the change in the relative momentum of the electrons, in the perpendicular direction. We have introduced the usual 'dimensionless intensity parameter' $\mu = eF/mc\omega$. The index $n$ of the Bessel function refers to the net number of photons (plasmons, in the present case), which are absorbed by the two-electron system. Our observations may indicate the same process, although not in vacuum, but in a neutral plasma. If we put into this formula the concrete values of our experiments and use a plasmon field enhancement factor of $g = 30$, then the intensity parameter becomes $\mu = geF_0/mc\omega$, where $F_0$ is the amplitude of the electric field strength of the incoming laser field. Fig. 7 shows the effective potential at 10 GW/cm², and Fig. 8 shows it for 120 GW/cm² laser intensity. In evaluating the $z_1$ values, we have expressed $(c\Delta p'_\perp/\hbar\omega) = \sqrt{(2mc^2/\hbar\omega)} \times \sqrt{\Delta E'_\perp/\hbar\omega}$, where the energy change $\Delta E'_\perp \equiv (\Delta p'_\perp)^2/2m$ of the relative motion has been taken $\Delta E'_\perp \approx kT$, i.e., the thermal energy. In the numerical illustrations we have taken $kT \approx 0.025 eV$, i.e. we have considered room temperature thermal energy. At low intensities the Bessel function is always positive, while at high ones negative regions can be observed. The existence of such regions may be responsible for the anomalies shown in Fig. 1 to Fig. 6.

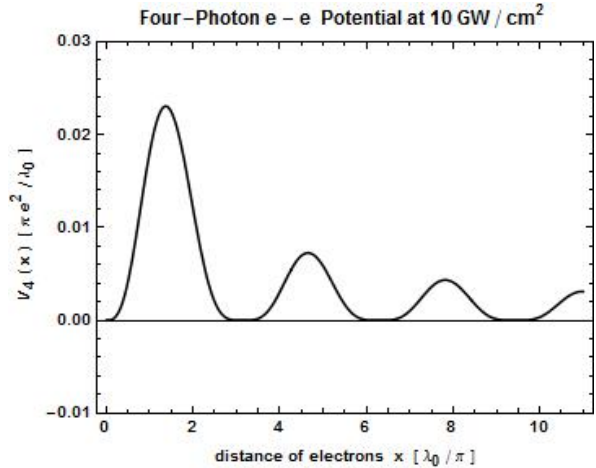

Fig. 7: Shows, on the basis of Eq. (1), the variation of the electron-electron effective potential along the propagation direction of the plasmon wave, in case of the four-photon absorption of the e-e pair ($n = 4$), for $I = 10 GW/cm^2$ incoming laser intensity. We have also taken into account the assumed field-enhancement factor $g = 30$, and $z_1 = 2$.

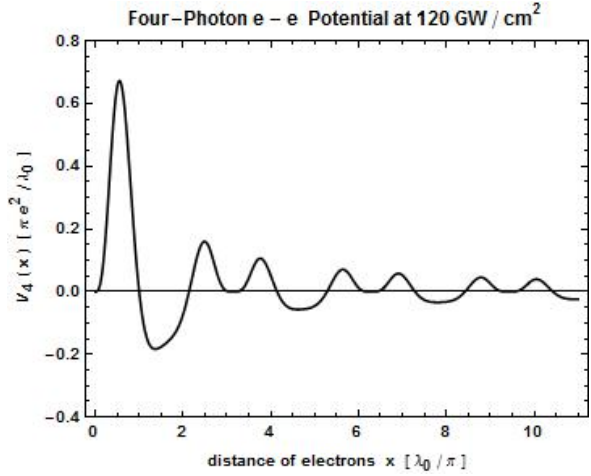

Fig. 8: Shows, on the basis of Eq. (1), the variation of the electron-electron effective potential along the propagation direction of the plasmon wave, in case of the four-photon absorption of the e-e pair ($n=4$) for $I=120 GW/cm^2$ incoming laser intensity, by assuming the same field-enhancement factor $g=30$ as in Fig. 7 but here $z_1 = 9$.

In our understanding the results of our two sets of present experiments and the qualitative interrelation of them with the effective attraction derived in [2] support the idea, that in strong electromagnetic fields electrons may be coupled into pairs at room temperature by surface plasmons on a gold surface at room temperature. These pairs may live for much longer times than the life-time of these plasmons or the time of the laser excitation. The electron emission is clearly due at least to two different processes, as indicated by the different slopes in the perturbative region of the two parts of the electron spectrum. The upper maximum may be due to the electron pairing effect and the lower one to 3 plasmon mediated emission.

Although we think that our experimental results offer an evidence for electron pairing, we are well aware of the fact that there are still several open questions, concerning the above qualitative interpretation, and further experimental and theoretical studies are to be carried out, concerning for instance the potential existence of long range phase coherence.

\*\*\*

We acknowledge support from the Hungarian Academy of Sciences. We wish to thank Peter Dombi and Győző Farkas for fruitful discussions.